\RequirePackage{lineno}
\documentclass[aps, prl,
floatfix,twocolumn,superscriptaddress,nofootinbib
]{revtex4-2}
\pdfoutput=1
\usepackage{t1enc}
\usepackage{gensymb}
\usepackage[utf8]{inputenc}
\usepackage[T1]{fontenc}
\usepackage{lmodern}
\usepackage{graphicx}
\usepackage{epsfig}
\usepackage{amssymb}
\usepackage{amsmath}
\usepackage{dcolumn}
\usepackage{bm}
\usepackage{color}
\usepackage{float}
\usepackage{hyperref}
\usepackage{natbib}
\usepackage{footnote}
\usepackage{ulem}
\usepackage{hhline}
\begin{document}

\title{Flat Bands at the Fermi Level in Unconventional Superconductor YFe$_2$Ge$_2$}

\author{R.\,Kurleto}
\affiliation{Department of Physics, University of Colorado, Boulder, CO, 80309, USA}
\author{S.\, Acharya}
\affiliation{National Renewable Energy Laboratories, Golden, CO 80401}
\author{C.-H.\, Wu}
\affiliation{Department of Physics, University of Colorado, Boulder, CO, 80309, USA}

\author{D.M.\, Narayan}
\author{B.S. Berggren}
\author{P.\,Hao}
\author{A.\,Shackelford}
\author{H.R.\,Whitelock}
\author{Z.\, Sierzega}
\affiliation{Department of Physics, University of Colorado, Boulder, CO, 80309, USA}

\author{M.\,Hashimoto}
\author{D.\,Lu}
\affiliation{Stanford Synchrotron Radiation Lightsource, SLAC National Accelerator Laboratory, Menlo Park, CA, 94025, USA}

\author{C.\,~Jozwiak}
\affiliation{Advanced Light Source, Lawrence Berkeley National Laboratory, Berkeley, California 94720, USA}

\author{R.P.\,Cline}
\affiliation{National Renewable Energy Laboratories, Golden, CO 80401}

\author{D.\, Pashov}
\affiliation{King’s College London, Theory and Simulation of Condensed Matter, The Strand, WC2R 2LS London, UK}

\author{J.\,Chen}
\affiliation{Cavendish Laboratory, University of Cambridge, Cambridge CB3 0HE, United Kingdom}

\author{M.\, van Schilfgaarde}
\affiliation{National Renewable Energy Laboratories, Golden, CO 80401}

\author{F.M.\,Grosche}
\affiliation{Cavendish Laboratory, University of Cambridge, Cambridge CB3 0HE, United Kingdom}

\author{D.S.\,Dessau}
\affiliation{Department of Physics, University of Colorado, Boulder, CO, 80309, USA}
\affiliation{Center for Experiments in Quantum Materials and RASEI, University of Colorado, Boulder, CO, 80309, USA}

\begin{abstract}\
We report heavy electron behavior in unconventional superconductor YFe$_2$Ge$_2$ ($T_C \,{=}\, 1.2$~K). We directly observe very heavy bands  ($m_\mathrm{eff}\sim 25 m_e$) within $\sim$10 meV of the Fermi level $E_{F}$ using angle-resolved photoelectron spectroscopy (ARPES). The flat bands reside at the X points of the Brillouin zone and are composed principally of $d_{xz}$ and $d_{yz}$ orbitals. We utilize many-body perturbative theory, GW, to calculate the electronic structure of this material, obtaining excellent agreement with the ARPES data with relatively  minor band renormalizations and band shifting required. We obtain further agreement at the Dynamical Mean Field Theory (DMFT) level, highlighting the 
emergence of the many-body physics at low energies (near $E_F$) and temperatures.


\end{abstract}


\maketitle

\section{Introduction}
Flat bands at the Fermi energy, or massive charge carriers, often gives rise to a rich variety of emergent states that can include novel superconductivity, density waves of various kinds, and even fractional quasiparticles.   
Flat bands can arise from electron hopping "matrix elements" that minimize the wavefunction overlap between neighboring atomic sites, or it can result from many-body effects where the many-particle electronic interactions conspire to further flatten or renormalize the electronic states. Importantly, both origins can be at play in the same material, with the first mechanism setting the conditions or environment that the many-body effects can then take advantage of and amplify.  The results we present here indicate that YFe$_2$Ge$_2$ is an especially clear and extreme version of such a situation. We observe bands of effective mass ($m_\mathrm{eff}\sim 25 m_e$) that are perhaps the heaviest bands directly observed in a d-electron system (f-electron systems that start from highly localized f-orbitals can have much heavier bands). Further, we find that the most massive portion of the bands are pinned to the Fermi energy within the experimental uncertainty of a few meV. This is a clear sign that electronic correlations are directly at play as such interactions have a tendency to drive the states to the Fermi energy where they also become the most electronically active. 

By comparing our results with theoretical calculations that include varying levels of sophistication of treating the interaction effects - Density Functional Theory (DFT) with the fewest interactions to Quasiparticle Self-Consistent GW (QSGW) with an intermediate level, to Dynamical Mean Field Theory (DMFT) with the most, we are able to unravel the physics that leads to the extremely flat bands at the Fermi level. Further, this program enable us to best understand the implications of these flat bands, for example for calculating the electronic susceptibility to various types of emergent states.

YFe$_2$Ge$_2$ shares the same body tetragonal structure with 122 iron pnictides (ThCr$_2$Si$_2$-type unit cell, I4/mmm space group). The transition to superconducting state is observed at temperature equal to~$1.2$~K in specific heat. The Sommerfeld coefficient, $\gamma\,{\approx}\,100$\,mJ/(mole K$^2$), is moderately enhanced which points to the nontrivial role of electron correlations. Proximity to the  putative quantum critical point (QCP) have been inferred by observing suppression of magnetic order (spin density wave state below 9~K in the parent compound LuFe$_2$Ge$_2$) in Lu$_{1-x}$Y$_x$Fe$_2$Ge$_2$ with increasing $x$. QCP is formed at $x\,{=}\,0.2$ and it is accompanied by a non-Fermi liquid behavior of electrical resistivity~\cite{sheng_YFG}. As noted, fluctuating magnetic moments on Fe ($\mu_\mathrm{Fe}\,{=}\,1\,\mu_B$) have been observed in the normal state at room temperature using angle-integrated photoemission spectroscopy and x-ray absorption spectroscopy~\cite{yfg_sirica}. Importance of electron-phonon coupling for the superconductivity was suggested on the basis of angle-resolved photoelectron spectroscopy (ARPES) results~\cite{arpes_yfg}. A kink was observed in the spectra collected along $\Gamma(\mathrm{Z}){-}X(\mathrm{R})$ direction at the binding energy $E_B\,{\approx}\,30$\,meV, which is close to the Debye temperature of YFe$_2$Ge$_2$ ($\Theta_D\,{\approx}\,403$\,K). Recent quantum oscillation measurements resolved bulk band structure of YFe$_2$Ge$_2$~\cite{malte}. Electronic structure of YFe$_2$Ge$_2$ has also been a subject of density functional theory studies, which predict magnetic ground state in contradiction with experiments~\cite{yfg_dft_1,yfg_dft_2}. Nuclear magnetic resonance (NMR) measurements on $^{89}$Y estimate the coherence temperature, $T^{\star}=75\pm15$~K~\cite{yfg_nmr}.

Here we report results of detailed studies of the electronic structure of YFe$_2$Ge$_2$ in the normal state. We have chosen angle-resolved photoelectron spectroscopy (ARPES) to observe band topography, because it is the most direct method~\cite{sobota_review}. It allows us to determine effective masses of charge carriers and discuss orbital composition of the bands, usually in connection with electronic structure calculations. We observed a dispersion related to heavy charge carries with an effective mass $m_\mathrm{eff}\sim 25m_e$ near the X point. This is the highest effective mass observed for iron-based superconductors, so far~\cite{malte}. Prior studies have shown  that the DFT calculations cannot describe the experimental band structure of YFe$_2$Ge$_2$ on a quantitative level. We use three different levels of the theory, DFT, many-body perturbative quasi-particle self-consistent Green's function (QS\textit{GW} ) approach and locally exact dynamical mean field theory (DMFT). The goal is to address both the origin of heavy mass and the Kondo physics and, in the process, disentangle the mechanisms that lead to the two.






\section{Methods}

Single crystals of YFe$_2$Ge$_2$ have been obtained by modified flux method. Details of sample preparation and characterization are described elsewhere~\cite{yfg_crystal}. 
Electronic structure of YFe$_2$Ge$_2$  has been studied using angle-resolved photoelectron spectroscopy (ARPES). Data presented in this paper have been collected at the BL5-2 beamline of SSRL synchrotron (California, US). Samples have been cleaved in situ in ARPES chamber (base pressure 10$^{-10}$ mbar) exposing (001) plane (conventional unit cell notation). Band mapping measurements have been performed at temperature equal to 14~K using Scienta DA30L photoelectron energy analyzer equipped with deflectors. Radiation with different polarizations (linear vertical, linear horizontal, circular right, and circular left) and energies between 25 and 170~eV has been used.  In all cases achieved energy resolution was better than 30~meV. 

We used experimental lattice parameters and atomic positions ($a\,{=}\,3.9617$~{\AA}, $c\,{=}\,10.421$~{\AA}) for the band structure calculations presented in this paper. Crystal structure optimization yielded atomic position and lattice constants in good agreement with those from previous x-ray diffraction experiment at room temperature~\cite{YFG_crystal_structure}. We study electronic structure of YFe$_2$Ge$_2$ at three different levels of theory: density functional theory (DFT) within the local density approximation (LDA), quasiparticle self-consistent \textit{GW}  approximation (QS\textit{GW})~\cite{QSGW_paper,questaal_paper}, and QS\textit{GW} combined with dynamical mean field theory (DMFT).~\cite{Sponza17,Acharya19} DFT calculations and energy band calculations with the static quasiparticle QS\textit{GW}  and self-energy $\Sigma^{0}(k)$ were performed on a 12$\times$12$\times$12 {\bf k}-mesh while the dynamical self-energy $\Sigma({\bf k})$ was constructed using a 6$\times$6$\times$6  {\bf k}-mesh and $\Sigma^{0}$({\bf k}) extracted from it.  For each iteration in the QS\textit{GW}  self-consistency cycle, the charge density was made self-consistent.  The QS\textit{GW}  cycles were iterated until the root-mean-square change in $\Sigma^{0}$ reached 10$^{-5}$\,Ry.  Thus the calculations were self-consistent in both $\Sigma^{0}({\bf k})$ and the density. DMFT calculations are performed using CTQMC~\cite{hauleqmc,gull} as the impurity solver. CT-QMC calculations are performed on five $d$-orbitals of Fe. While performing nonmagnetic QS\textit{GW} + paramagnetic DMFT, the Hubbard model in DMFT is solved using $U\,{=}\,0.12$\,eV, $J\,{=}\,0.024$\,eV and double counting correction as implemented in the fully localized limit,  E$_{DC}\,{=}\,0.6$\,eV. Within DMFT, calculations were performed in the temperature window of 2000\,K to 20\,K to explore the lattice incoherence-coherence crossover where the static part of the dynamic spin susceptibility shows a change from high-temperature Curie-Weiss ($\frac{1}{T}$) to low-temperature Pauli behavior (T$^{0}$).

\section{Results}

Knowledge of band structure is important for understanding properties of superconducting materials. The superconducting gap can vary with wave number and can have non-trivial band dependence~\cite{Ding_2008}. Also, more exotic behaviors, like multiple transitions to a superconducting state as a function of temperature (presence of more than one critical temperatures) can be explained by band structure effects. In the last case, superconducting gaps open at different temperature for different bands crossing the Fermi level. Motivated by this, we have performed detailed exploration of band topography of the unconventional superconductor YFe$_2$Ge$_2$. Fermi maps collected with ARPES at low temperature (14~K) are shown in~Fig.~\ref{fig:1}. They deliver additional information with respect to previous ARPES study by Xu et al~\cite{arpes_yfg}. That previous work delivered important information about band structure and experimental evidence hinting on importance of electron-phonon coupling for the superconductivity in the material.  On the other hand, a consensus on matching bands in three-dimensional BZ has not been reached, so far. Xu et al., matched the renormalized DFT band structure in $\Gamma$-$\Sigma$-$X$ plane, with mass enhancement factors not exceeding~3~\cite{arpes_yfg}. The agreement in $Z$-$\Sigma$-$Y$ plane was not satisfactory, even on a qualitative level.

Significant electronic correlations in iron based superconductors have been evidenced by multiple experimental techniques. Proper explanation of their effects on physical properties calls for use of advanced many body techniques. Density functional theory is a good starting point in many cases. However, it is common that the theoretical band structure needs to be adjusted to account for mass renormalization.  Band dependent scaling factors and shifts are usually introduced, which makes unique match with experimental band structure difficult. In consequence, the predictive power of DFT calculations is weakened. This issue can be overcome by including (partially, beyond mean-field level) electronic correlations, which in principle, can be separated in to local and non-local part. The first contribution is described well by the dynamical mean-field theory (DMFT), while the second is captured in the quasiparticle self-consistent GW (QS\textit{GW} ) method~\cite{mark_prl}. QS\textit{GW}  is a single-particle theory that incorporates the long-range nature of the charge-correlations in a diagrammatic fashion, while DMFT can incorporate local spin-fluctuations in an exact manner.

\subsection{Band structure topography /  Fermiology}
\begin{figure*}
\centering\includegraphics[width=0.99\linewidth]{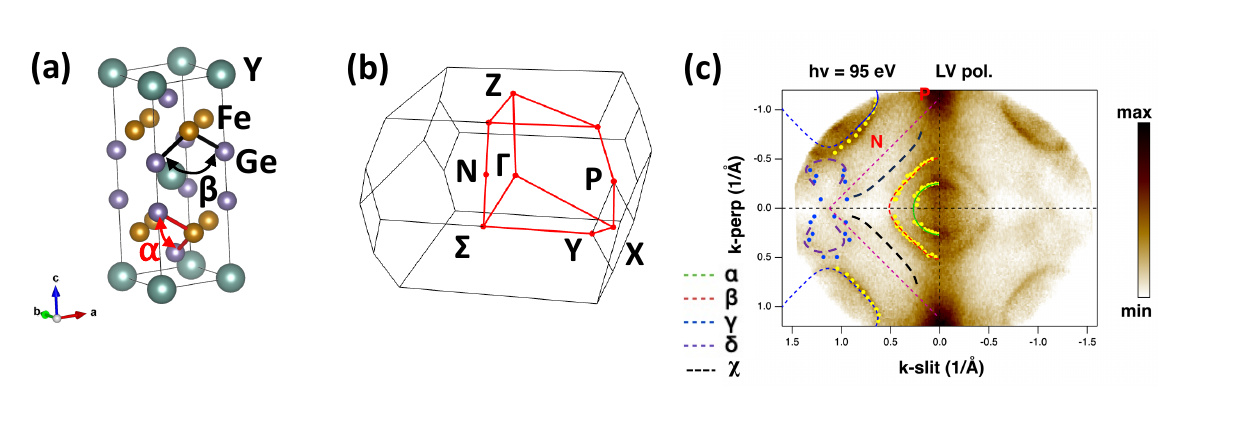}
\caption{Results of ARPES measurements performed on YFe$_2$Ge$_2$ at low temperature ($T = 14$~K). Unit cell of YFe$_2$Ge$_2$ is shown in panel~(a) with Fe-Ge angles marked as $\alpha$ and $\beta$. The  first Brillouin zone is shown in panel~(b) (we use labeling convention consistent with Ref.~\cite{bz_bct}). (c)~Fermi maps measured with Linear Vertical (LV) polarization corresponding to the $P-N$~plane in BZ ($h\nu = 95$~eV). Linear Horizontal (LH) polarization (not shown) helps highlight certain of the contours more strongly.  Extracted Fermi surface contours are superimposed on the left half of the color map ($\alpha$, $\beta$, $\gamma$, $\delta$, $\chi$).}
\label{fig:1}
\end{figure*}

The crystal structure of YFe$_2$Ge$_2$ is shown in~Fig.~\ref{fig:1}~a, where we marked characteristic Fe-Ge angles, showing significant deviation from perfect tetrahedral Fe environment, which is almost realized for optimally doped K$_{0.4}$Ba$_{0.6}$Fe$_2$As$_2$. Our systematic ARPES measurements allowed us to determine photon energies corresponding to high symmetry points in three dimensional Brillouin zone~(BZ) (b): $h\nu = 59$, 160~eV corresponds to $Z$ planes, $h\nu=95$, 152~eV corresponds to $P$-$N$ planes, and $h\nu=81$, 138~eV corresponds to $\Gamma$ planes. 
 We present the fermiology of YFe$_2$Ge$_2$ in~Fig.~\ref{fig:1}~c. At least five different contours have been identified on the Fermi surface: $\alpha$,~$\beta$,~$\gamma$,~$\delta$,~$\chi$. The inner circular contour around the BZ center, marked as $\alpha$, corresponds to a hole pocket characterized by a strong $k_z$ dispersion. The outer contour, marked as $\beta$, can be described by rounded square shape and is also related to holelike band. However, its dispersion along $k_z$ direction is relatively weak. 
This band is visible in adjacent Brillouin zones as well,  but we are using distinct label $\gamma$ for clarity. Interestingly, the small pockets $\delta$ are found around the zone corners ($\Gamma$-$X$ or $Z$-$Y$ direction). We describe them with a propeller type contour, but we note that the interpretation of these contours is challenging. In addition, we note that several shallow pockets with different curvatures cross $E_F$ around zone corners. Other contours are also visible in presented Fermi maps  (for example $\chi$), but their interpretation is not straightforward. We also measured maps at photon energies corresponding to the $\Gamma$  point at the normal emission. The features present there (not shown) can be interpreted as a fourfold starpshape contours closed around the $\Gamma$ point or alternatively as a two-fold shape closed around the $\Sigma$ point. To make a final conclusion about topography of these FS sections we would need to connect our ARPES results with those from bulk sensitive quantum oscillation measurements. This is why we limit ourself to presentation of sections of FS at $P-N$ plane, as these data can be interpreted without ambiguity.

\subsection{QS\textit{GW}  calculations match to ARPES spectra}

\begin{figure*}
\centering\includegraphics[width=0.99\linewidth]{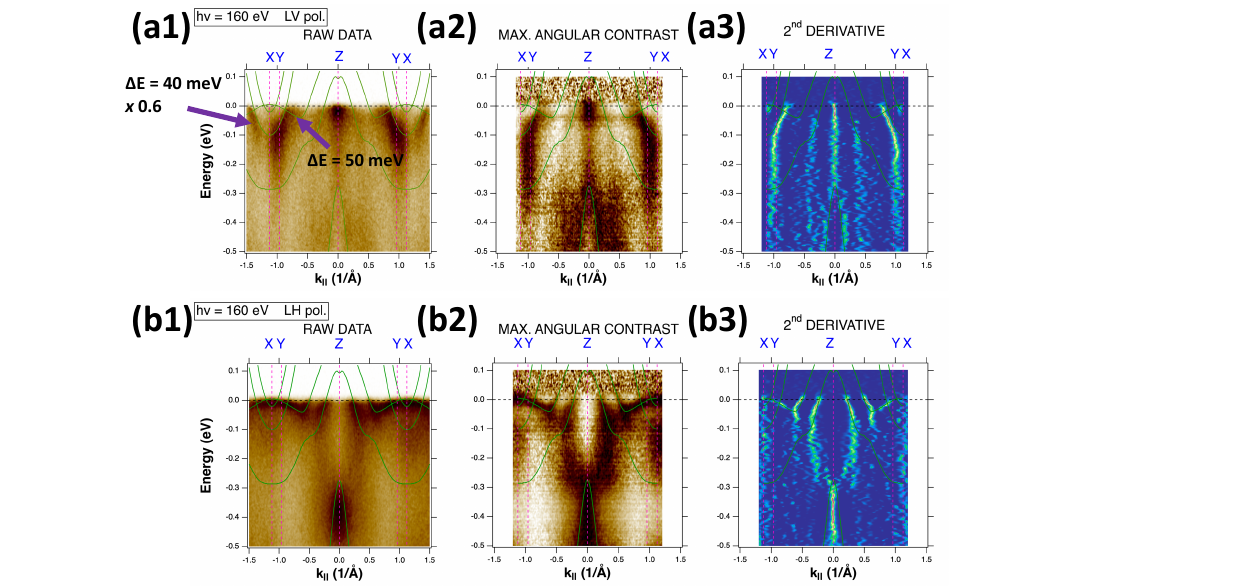}
\caption{QS\textit{GW}  calculations match with ARPES spectra along $X-Y-Z$ direction. Experimental data were collected at $T=14$~K with $h\nu=160$~eV using LV (a1-a3) and LH (b1-b3) polarization. To enhance the data contrast we are showing results of maximal angular contrast (MAC) procedure (a2, b2) and second derivative (a3,b3) applied to original data (a1,b1). Some bands have been renormalized and shifted as indicated in panel (a1). Other bands match these spectra without any adjustment.}
\label{fig:3_supp}
\end{figure*}

Fig.~\ref{fig:3_supp} shows the match of QS\textit{GW}  calculations to band structure measured with ARPES. We analyzed the relation between theoretical bands dispersion for many different photon energies, but we present only results for $h\nu=160$~eV, as this data is of the most interest. We reached the excellent agreement between calculated bands and experimental data close to the Fermi level after introducing minor adjustments. The outer hole pocket around Z point connects with flat band around X points. They need to be shifted rigidly by 50~meV to agree with experimental spectrum for $Z-X$ path collected with $h\nu=160$~eV. Electron pocket at X point needs renormalization of factor 0.6 and 40 meV upward shift.  Other bands seem to agree reasonably well between $E_B\,{=}-100$~meV and Fermi level without any adjustments. At deeper binding energies we can see the discrepancy between the continuation of the inner hole pocket which should form a wide parabolic electron pocket around the X point with a bottom at $-0.3$~eV. Such a band is not observed in experimental data.

\subsection{Flat band}

In addition to bands in $P-N$ plane we would like to focus on band dispersions in $Z-\Sigma-Y$ plane (this plane connects with $\Gamma - X -\Sigma $ section of BZ, as a consequence of bct structure). Bands observed in $P-N$ plane were relatively steep, with effective masses not exceeding $10m_e$ (for $\beta$ band we got $m_{eff}\approx5m_e$ from the fit between to -100~meV and $E_F$, but QS$GW$ still seems to match very well on extended energy scale). The data collected at $h\nu=160$~eV, i.e., at $k_z$ corresponding to the $Z$ plane in the 3D BZ show different behavior (Fig. 2). Fermi surface topology just around the $Z$ point is analogous to the one in $P-N$ plane, with two holelike contours in the center.  Again, we do not discuss particular shape of bands visible around the $\Gamma$ points in adjacent zones as unique description is not possible with the present data (they may be described as star/propeller shaped single closed contours or several disjoint oval pockets). Other new features are hot spots of high intensity blurred around zone corners, i.e., along $Y-X-Y$ segments interconnecting adjacent zones. These are solely visible in spectra collected with LH polarization and only along the slit direction (horizontal direction in the plot). This type of selectivity can already point to the significant orbital polarization in the band structure. The spectra measured along $Z-Y-X$ direction show dispersions related to features identified in the FS map. We can see that the steep bands around the Z point coexist with the very flat bands around X points. The first ones are marked with blue and red lines, and they are characterized by effective masses equal to $2.9m_e$ and $13m_e$, respectively. The values are obtained from parabolic fits to dispersion obtained from Lorentzian fits to momentum distribution curves (MDC). The QS\textit{GW} effective masses are respectively 2.0$m_e$ and $4.2m_e$. The experimental hole bands at the X points (marked with green lines) have effective mass approximately equal to $25m_e$. This value comes from parabolic fits to energy distribution curves (EDC) derived dispersion and it is a reasonable estimation, because similar (or higher) values have been obtained from our other analysis methods (2D spectral function fits - not shown). On the contrary, LV spectrum shows the electron pockets around the X points. To better visualize dispersive feature near the Fermi level we also show results of maximum angular contrast (MAC) procedure in panels c and e. The QS\textit{GW} effective mass is $19.0m_e$, but this grows to $24m_e$ in our DMFT calculations, matching the ARPES data well.

\begin{figure*}
\centering\includegraphics[width=0.99\linewidth]{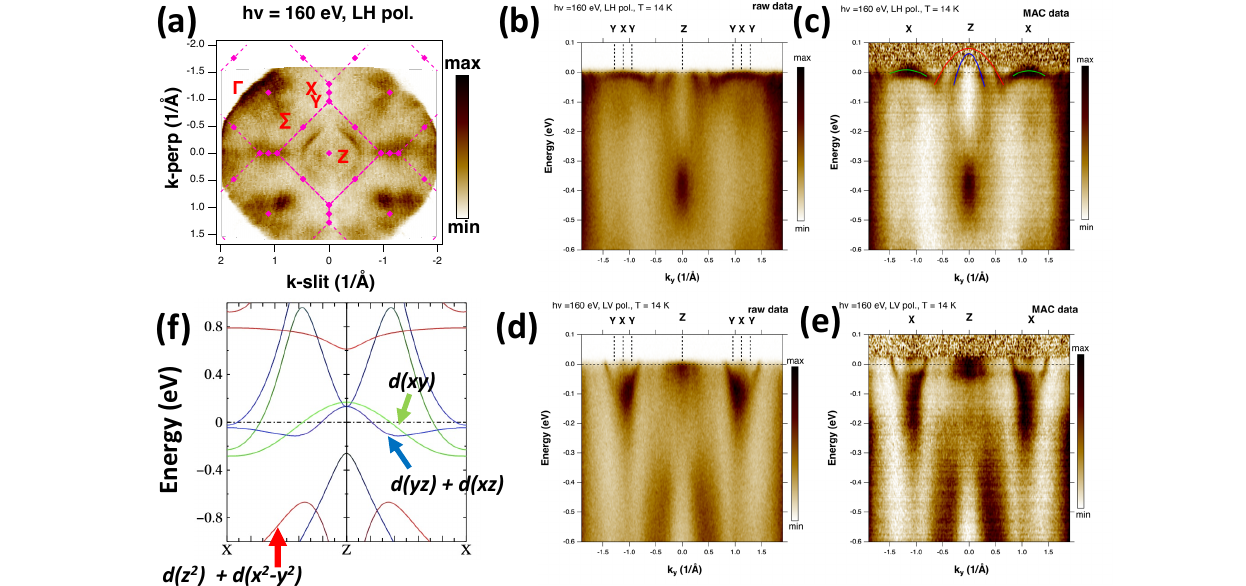}
\caption{Flat bands in electronic structure of YFe$_2$Ge$_2$. (a) Fermi map measured at $h\nu=160$~eV, what corresponds to $Z$ point at the normal emission. Hot spots of intensity related to flat bands are visible along $Y-X-Y$ lines for $k-perp = 0$. High resolution spectra measured along $Z-X$ direction with (b) LH  and (d) LV polarization. Results of maximum angular contrast procedure~\cite{mac} are shown in panels  (c) and (e), respectively. (f) QS\textit{GW}  calculations along $Z-Y$ direction with marked leading orbital characters. Germanium states are marked with black color.}
\label{fig:2}
\end{figure*}


We observe systematic changes in the band energies and band renormalizations going from DFT to QS\textit{GW}  to DMFT. The electronlike and the flat holelike bands at $X$ point are respectively 70 and 90 meV below E$_{F}$ in DFT and only 15 and 40 meV below E$_{F}$ in QS\textit{GW}. Most importantly, the non-local self-energy in QS\textit{GW}  leads to a factor of two enhancement in the band-mass of the flat d$_{xz,yz}$ like band at the $X$ point compared to DFT. This happens primarily because the out-of-plane component of the mass tensor gets heavier by a factor of two owing to the GW self-energy corrections. This is a crucial observation considering that QS\textit{GW}  neither invokes higher order magnetic scattering nor the Hund's physics, the way it is generally perceived. The simple fact that a non-magnetic many-body perturbative theory gets most of the band renormalization as observed in ARPES is indicative that it is primarily a one-particle effect, akin to the emergence of flat band in twisted-bilayer graphene. We will elaborate on this in a later section.

Another insight into the electronic structure of YFe$_2$Ge$_2$ is orbital content of particular bands. According to the calculations, states near $E_F$ originates mainly from Fe $3d$ orbitals. The exception is the top of the band visible at $E_B\,{\sim}-0.4$\,eV, which is of Ge character (marked by black line, see~Fig.~\ref{fig:2}).  Both the outer hole pocket around the Z point and the electron pocket around the X point are dominated by Fe $d(xy)$ orbitals (indicated by green color). The heavy hole pocket around the X point and the inner pocket around the $Z$ point are connected within one band and they are combinations of Fe $d(yz)$ and $d(xz)$ orbitals (blue line). Other Fe orbitals, i.e., $d(z^2)$ and $d(x^2 - y^2)$, are shifted away from $E_F$, either $\approx0.8$~eV above or below $E_F$ (indicated by red). According to selection rules for photoemission matrix elements in our experimental geometry  LH polarized radiation should probe states which are related to antisymmetric combinations of $d(xz)$ and $d(yz)$ orbitals polarization together with $d(xy)$ and $d(z^2)$ states in $\Gamma - X$ orientation. States with $d(x^2-y^2)$ or symmetric combination of  $d(xz)$ and $d(yz)$ orbitals should be detected with LV polarization in this orientation.


Strong orbital differentiation in both one-~\cite{orbitalselective1,orbitalselective2} and two-particle sectors ~\cite{acharya2023vertex} are some of the main features identifying the Hund's metals, however, as we show below, the situation is quite distinct in YFe$_{2}$Ge$_{2}$. We observe that the Fe-3$d$ occupancy is 5.98 (d$_{x^2-y^2}$ and d$_{xy}$ with 1.177 electrons each, d$_{xz}$ and d$_{yz}$ with 1.21 electrons each and d$_{z^2}$ with 1.204 electrons) in DFT, 5.82 (d$_{x^2-y^2}$ and d$_{xy}$ with 1.144 electrons each, d$_{xz}$ and d$_{yz}$ with 1.178 electrons each and d$_{z^2}$ with 1.174 electrons)  in QS\textit{GW}  and 5.83 in DMFT. This is in contradiction to the atomic valence analysis which suggests that Fe could have a 3$d^{5.5}$ valence state. This is primarily because the material has significant covalent character and an ionic analysis does not give the correct valence in such cases. This also reflects in the fact that the low-energy magnetic susceptibility peaks at the ferromagnetic vector~\cite{wo2019coexistence} unlike many other iron based superconductors which are more ionic in nature and where the peak appears only at the antiferromagnetic vector in the Fe-square-plane.  The Ge height over the Fe plane is 1.33~{\AA}, which is closer to the As height (1.357~{\AA}) in the uncollapsed tetragonal phase~\cite{Acharyalafeas} of LaFe$_{2}$As$_{2}$ and much shorter than the Se height in FeSe (1.463~{\AA})~\cite{yi2017role}. The d$_{xz,yz}$ band renormalization as observed in ARPES compared to DFT correlates well with the established phenomenological curves showing d$_{yz}$ renormalization against Fe-pnictogen/chalcogen bond length and Fe-3$d$ filling for a large class of iron based superconductors.~\cite{yi2017role,lee2008effect,mizuguchi2010anion}. The Fe-Ge-Fe bond angle is 111.72$\degree$  which is similar to weakly Co or K doped BaFe$_{2}$As$_{2}$ and is much larger than Fe-chalcogen-Fe bond angles from more ionic candidates like FeSe and FeTe. 

\begin{figure*}
\centering\includegraphics[width=0.6\linewidth]{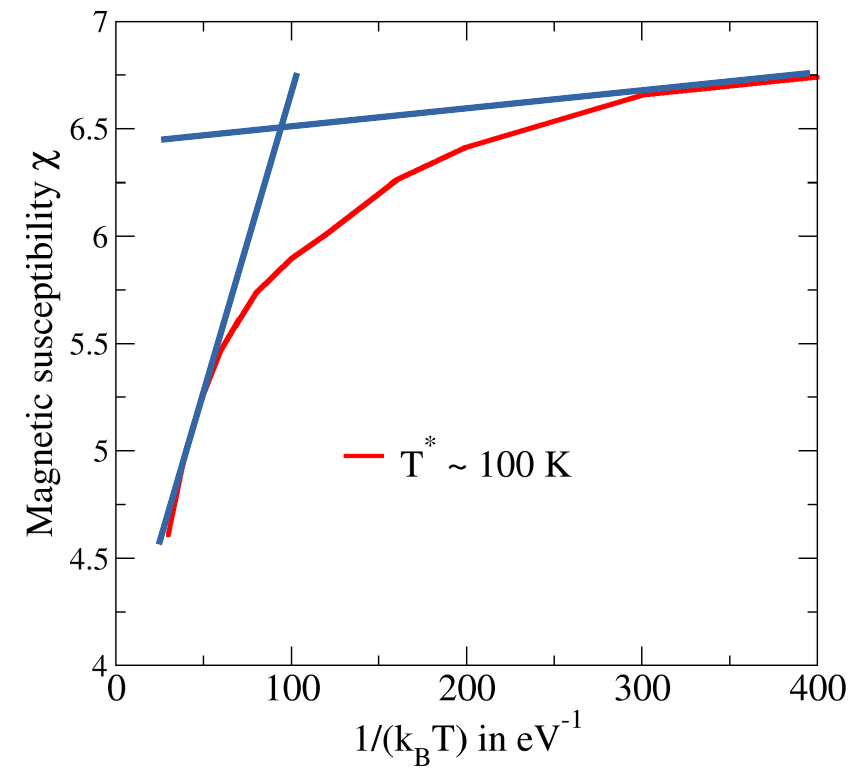}
\caption{Temperature dependent incoherence-coherence crossover in local magnetic susceptibilities $\chi$ (in arbitrary units) computed from DMFT. The susceptibility changes from high temperature ($1/(k_B T)$ below $100$~eV$^{-1}$, that is $T \approx116$~K) linear in $\frac{1}{T}$ behavior to a nearly temperature independent behavior at  low temperatures (towards $1/(k_B T)\approx400$~eV$^{-1}$, that is $T \approx30$~K). The coherence temperature is denoted by $T^{\star}\sim100$~K.}
\label{fig:5}
\end{figure*}

All of these observations are consistent with the fact that the system does not have large spin correlations and a QS\textit{GW}  theory that incorporates the long-range nature of the charge-correlations in a diagrammatic fashion, already describes most of the electronic properties of the material sufficiently. To put it in perspective, in FeSe, spin fluctuations characterized by $\langle{S^2}\rangle\,{=}\,5\,\mu^{2}_{B}$~\cite{gretarsson2011revealing} (and DMFT plays an important role~\cite{yin2014spin,acharya2022role}) while in YFe$_{2}$Ge$_{2}$ it is 1~$\mu^{2}_{B}$.  Further weak corrections are required on top of QS\textit{GW}  to produce almost exact agreement with ARPES and that is primarily because the QS\textit{GW}  calculations are non-magnetic in nature and the real material is paramagnetic. This is why small corrections like $U$=0.12 eV and $J$=0.024 eV within DMFT (that incorporates local spin fluctuations in a paramagnetic environment missing from QS\textit{GW}) are sufficient to produce almost exact agreement with ARPES data at low energies. However, once the flat band with heavy electron-mass appears at the Fermi energy it has significant impact on the temperature dependent evolution of the electronic scattering. This is a physics which is caught well within DMFT which incorporates the temperature dependence of the local spin fluctuations with an impurity solver like CTQMC.
 The most definitive signature for the Kondo physics is the crossover in magnetic (local) susceptibility that takes it from a high temperature Curie-Weiss to a low-temperature Pauli susceptibility~\cite{Haule_2009,mravlje2011coherence}. Such a crossover is often a clear signature for the lattice coherence scale setting in. A similar incoherence-coherence crossover is observed in Hund's metal Sr$_{2}$RuO$_{4}$ too~\cite{sidis1999evidence} and is well characterized by DMFT~\cite{acharya2017first}. We see a similar crossover (see Fig.~\ref{fig:5}) from our temperature dependent magnetic susceptibilities computed with DMFT for YFe$_{2}$Ge$_{2}$ around 100~K.

\section{Discussion}


We notice that the situation in YFe$_2$Ge$_2$ is partially analogous to LiV$_2$O$_4$~\cite{shimoyamada_2006}, which is considered as a $d$-electron heavy fermion system. The mass renormalization there appears due to the hybridization between two bands (with bandwidths 1 eV and 2 eV respectively), with strong Coulomb interactions leading to partial electron localization, despite the lack of drastic differences in bandwidths that are usually encountered in $f$-electron systems. Similar heavy fermion behavior in $d$-electron system has been also observed  for CaCu$_3$Ru$_4$O$_{12}$~\cite{liu_102}. However, the conclusion on the origin of the flat band close to $E_F$ was not in the favor of the Kondo interaction in that case. 

Here, we show that the origin of the flat band mass is not primarily down to Hund's physics, as is understood generally for the normal phases of some iron based superconductors. YFe$_{2}$Ge$_{2}$ has a significant covalent character and the heavy and light band masses can be described from one single level of theory that takes into account covalency and not higher order spin scattering that is characteristic of Hund's physics.  Once the flat band arrives at the Fermi energy it leads to many-body correlations and of particular interest is the Kondo physics. The main evidence for the validity of this Kondo scenario is our temperature dependent
local magnetic susceptibilities that shows a Curie-Wess behavior at high temperatures and nearly temperature independent susceptibilities at low temperatures (Fig.~\ref{fig:5}). 

The sharp and flat band we observe at $E_F$ thus may be considered to be a so-called Kondo resonance peak, with our preliminary temperature dependent ARPES data (not shown) supporting this possibility. In such a scenario the sharp peak exists below some characteristic temperature of the system (Kondo temperature $T_K$ or coherence temperature $T^{\star}$), which would also be consistent with bulk transport and specific heat measurements. 

YFe$_2$Ge$_2$ is clearly a very interesting electronic system with many exotic properties, including the superconductivity that may be on the verge of spin-triplet pairing~\cite{yfg_dft_1,yfg_nmr}, and a power low behavior of electrical resistivity as a function of temperature. 
The flat band that we observe may be especially relevant for understanding these exotic properties.

\section{ ACKNOWLEDGMENTS}
Work at CU-Boulder was funded by the U.S. DOE, Office of Science, Office of Basic Energy Sciences under Award Number DE-FG02-03ER46066, by the Gordon and Betty Moore Foundation's EPIQS initiative through grant no. GBMF9458, and by a seed grant from RASEI. Use of the Stanford Synchrotron Radiation Lightsource, SLAC National Accelerator Laboratory, is supported by the US DOE, Office of Science, Office of Basic Energy Sciences under Contract No. DE-AC02-76SF00515. We are grateful to Jonathan Denlinger (ALS) for his support during ARPES measurements. The theoretical part of the work was authored by the National Renewable Energy Laboratory, operated by Alliance for Sustainable Energy, LLC,
for the U.S. Department of Energy (DOE) under Contract No. DE-AC36-08GO28308, funding from Office of Science, Basic
Energy Sciences, Division of Materials. SA, MvS, DP, RPC acknowledge the use of the National Energy Research Scientific Computing Center, under Contract
No. DE-AC02-05CH11231 using NERSC award BES-ERCAP0021783 and we also acknowledge that a portion of the research was performed using the Eagle facility sponsored by the Department of Energy's Office of Energy Efficiency and Renewable Energy and located at the National Renewable Energy Laboratory. The work was also supported by the EPSRC of the UK (grants no. EP/K012894 and EP/P023290/1).

\bibliographystyle{apsrev4-1}
\bibliography{ARPESmany,theory}

\clearpage

\section{Supplementary material}

\subsection{Comparison of DFT and QS\textit{GW}  calculations}

\begin{figure*}
\centering\includegraphics[width=0.99\linewidth]{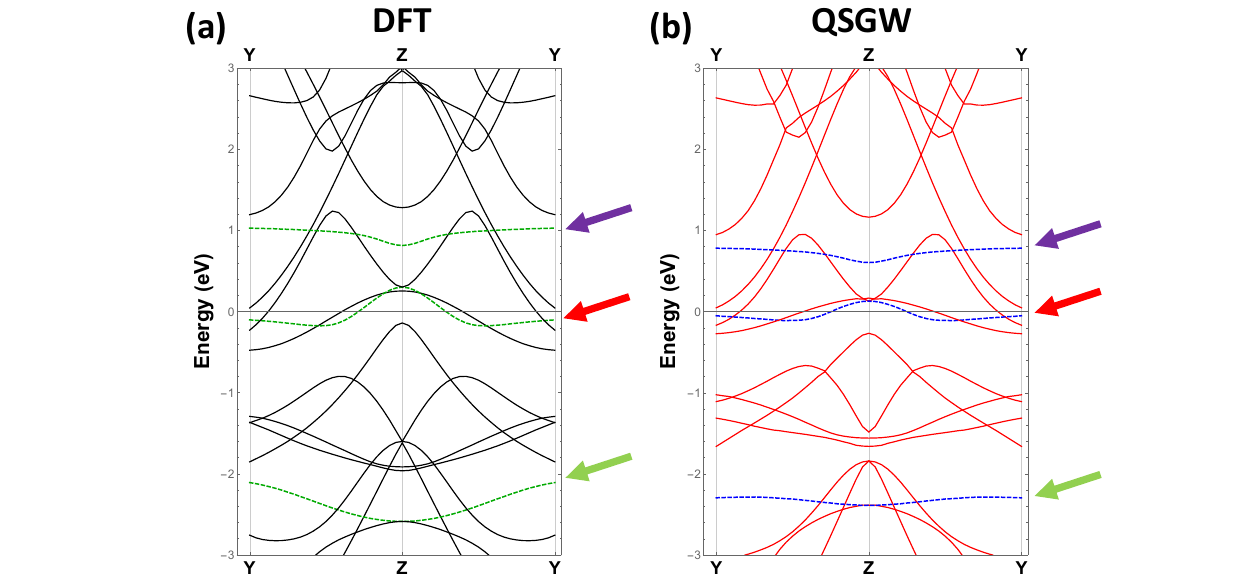}
\caption{Comparison between DFT (a) and QS\textit{GW}  (b) calculations with no spin-orbit coupling included.  Flat bands identified in both types of calculations are marked with dashed lines. Arrows of the same color points to the corresponding bands in different calculations.}
\label{fig:2_supp}
\end{figure*}

In the main part of the manuscript we state that we are not able to describe properly the band structure of YFe$_2$Ge$_2$ using DFT calculations. Though, satisfactory agreement is reached when QS\textit{GW}  method is employed. Here, we are demonstrating that some essential features of electronic structure are already present at DFT level. Fig.~\ref{fig:2_supp} compares two computational approaches: DFT (panel a) and QS\textit{GW}  (panel b). In both cases spin-orbit coupling was not included. This is because the spin-orbit coupling may produces anticrossings which depends on the details of band hybridization. Introduced complexity of the band structure in this way, might defeat the purpose of this simple analysis. We also note that the exact degeneracy of some bands at Z point is an artifact due to lack of spin-orbit coupling in the calculations. We were able to identify the sequence of flat bands in QS\textit{GW}  calculations (marked by blue dashed lines in panel b). The lowest of them is derived from Ge states, while the others are almost solely Fe $3d$ states. Interestingly, the bands with similar shapes can be identified in DFT calculations (marked by green dashed lines in panel a). We were able to connect DFT bands with their QS\textit{GW}  counterparts - the pairs of corresponding bands are indicated with the arrows of the same colors. The main result of correlations effects included on QS\textit{GW}  level is a significant reduction of the band widths and in some cases shift on the binding energy scale. In some cases gaps/anticrossings open around the points of degeneracy of DFT bands. This is particularly visible at Z point at $E_B\approx -1.6$~eV, as at least triple degeneracy at DFT level, which is lifted at QS\textit{GW}  level. Concluding, the DFT calculations seem to reflect the shape band structure of the material in general. However, to properly describe the features on the small energy scales (of the order of tens of meV) one needs to employ more realistic approach, as for example QS\textit{GW} .

\end{document}